# Superconductivity with valley-imbalanced Fermi surfaces in thick rhombohedral graphene

Le Liu[1#], Jian Zhao[1#], Cheng Xu[2#], Kenji Watanabe[3], Takashi Taniguchi[4], Yang Zhang[5,6*], Weibo Gao[1,7,8,9*]

[1] *School of Electrical and Electronic Engineering, Nanyang Technological University, Singapore 639798, Singapore.*

[2] *Max Planck Institute for Chemical Physics of Solids, 01187, Dresden, Germany*

[3] *Research Center for Electronic and Optical Materials, National Institute for Materials Science, 1-1 Namiki, Tsukuba 305-0044, Japan*

[4] *Research Center for Materials Nanoarchitectonics, National Institute for Materials Science, 1-1 Namiki, Tsukuba 305-0044, Japan*

[5] *Department of Physics, National University of Singapore, 117551, Singapore*

[6] *Department of Physics and Astronomy, University of Tennessee, Knoxville, Tennessee 37996, USA*

[7] *Division of Physics and Applied Physics, School of Physical and Mathematical Sciences, Nanyang Technological University, Singapore 637371, Singapore*

[8] *Centre for Quantum Technologies, National University of Singapore, Singapore, Singapore*

[9] *Quantum Science and Engineering Centre (QSec), Nanyang Technological University, Singapore, Singapore*

# These authors contributed equally

* Corresponding authors. Email: yangzhang@nus.edu.sg; wbgao@ntu.edu.sg

## ABSTRACT

Superconductivity in spin-polarized metals challenges conventional opposite-spin pairing, particularly when a magnetic field strengthens rather than suppresses it. In graphene, the additional valley degree of freedom broadens the possible pairing channels, making the valley occupation of the parent metal central to understanding such superconductivity. Here we report in-plane-field-enhanced superconductivity in hBN-aligned 17-layer rhombohedral graphene. Normal-state quantum oscillations reveal unequal Fermi pockets consistent with a valley-imbalanced, spin-polarized parent metal. Calculations show that the parent electron pocket is predominantly localized on the bottom surface and largely preserved by electrostatic screening as additional Fermi pockets emerge. Superconductivity persists in this regime but is suppressed when carriers are polarized toward the hBN-aligned surface, where the parent bands undergo strong moiré reconstruction. An adjacent nonlinear metallic response evolves continuously into superconducting critical-current features and persists above the superconducting transition temperature. The combination of valley imbalance and in-plane-field-enhanced superconductivity motivates a phenomenological magnon-assisted equal-spin intervalley pairing scenario. These findings provide experimental constraints on pairing in valley-imbalanced metals and open a route to testing magnetic pairing mechanisms through layer-selective control of the parent electronic states.

## I. INTRODUCTION

Rhombohedral graphene hosts surface-layer-polarized low-energy bands with a large density of states and nontrivial topological properties near charge neutrality [1,2]. This unique band structure has enabled a variety of interaction-driven symmetry-breaking metallic phases [3–6], and further supports superconductivity [7–28], magnetism [3,29] and topological states [30,31]. More recently, signatures of chiral superconductivity have been reported in rhombohedral tetra- and penta-layer graphene [11], emerging from a quarter-metal phase with spin and valley polarization. Separately, introducing a moiré potential by aligning rhombohedral graphene with hBN produces a sequence of correlated gaps at commensurate moiré fillings [32–37]. Intriguingly, upon hBN alignment, the region that otherwise supports chiral superconductivity instead gives rise to integer and fractional quantum anomalous Hall states [34,35,38]. The quasiparticles associated with these phases, including possible Majorana fermions and anyons, are of considerable interest for topological quantum computation [39,40].

Across these systems, superconductivity appears closely connected to the topology and flavor character of the parent Fermi surface [7,8,10–13,20]. Superconducting states have been observed in regimes with annular Fermi surfaces [7], near isospin-symmetry-breaking transitions [7,11,14,17,22,24], and in spin-orbit-proximitized regimes with spin-valley locking [9,12,13,41]. Spin polarization constrains pairing toward equal-spin triplet channels, while valley polarization or imbalance determines whether pairing is intra- or intervalley [42,43]. A mismatch between the paired Fermi surfaces is precisely the condition under which finite-momentum, Fulde–Ferrell–Larkin–Ovchinnikov (FFLO)-like pairing arises [44]. Resolving the parent Fermi surface is therefore essential for constraining the spin and momentum channels available for pairing.

A particularly instructive precedent is provided by ferromagnetic superconductors such as $UGe_2$, URhGe and UCoGe, where superconductivity develops within a spin-polarized metal and is widely associated with equal-spin triplet pairing mediated by ferromagnetic fluctuations rather than phonons [45–52]. In these and related systems, magnetic fields can even revive or strongly enhance superconductivity [51,53]. These materials, however, are largely fixed by composition and pressure. Rhombohedral graphene offers an electrostatically tunable counterpart, in which the spin–valley structure of the superconducting parent state can be characterized directly.

Here we investigate in-plane-field-enhanced superconductivity in hBN-aligned 17-layer rhombohedral graphene. In this thick-layer regime [54], self-consistent Hartree calculations show that screening by additional carriers largely preserves the bottom-surface electron pocket along the superconducting region. Quantum oscillations reveal unequal K and K' pockets consistent with a valley-imbalanced, spin-polarized parent metal. With decreasing displacement field, the parent state evolves from full spin and valley polarization to this imbalanced metal. An adjacent nonlinear metallic response evolves continuously into superconducting critical-current features and persists above the superconducting transition temperature. Superconductivity survives as additional pockets emerge but is suppressed when carriers are polarized toward the hBN-aligned surface, where strong moiré reconstruction occurs. The valley-imbalanced parent state and in-plane-field-enhanced superconductivity motivate a phenomenological magnon-assisted equal-spin intervalley pairing scenario, which we discuss alongside alternative mechanisms.

## II. RESULTS

### A. In-plane-magnetic-field-enhanced superconductivity near a spin-polarized phase boundary

Figure 1a shows a schematic of the device, consisting of a 17-layer rhombohedral graphene flake encapsulated between hBN flakes and aligned with the top hBN. The rhombohedral stacking order is identified by infrared imaging and Raman spectroscopy, and the thickness is determined by atomic force microscopy (see Fig. S1 in Supplementary Note 1). The dual-gate geometry allows independent control of the carrier density ($n$) and displacement field ($D$). Figure 1b shows an $n$-$D$ map of the four-terminal longitudinal resistance $R_{xx}$, where resistance peaks at integer moiré fillings are associated with correlated gaps and the moiré band gap. The extracted full-filling carrier density, $n_0 = 4.2 \times 10^{12}/\mathrm{cm}^2$, corresponds to a twist angle of 0.98° and a moiré period of approximately 10.4 nm. On the electron-doped side, positive $D$ polarizes carriers toward the bottom surface, away from the aligned hBN interface (weak-moiré side), whereas negative $D$ polarizes carriers toward the top surface (strong-moiré side). Consistent with this layer-polarization picture, the calculated band structure of 17-layer rhombohedral graphene at $D = 0.7$ V/nm (Fig. 1c) exhibits a nearly flat conduction-band edge, with the corresponding low-energy states predominantly localized on the bottom surface.

Within the region outlined by the white dashed lines in Fig. 1b, a narrow resistance feature emerges near moiré filling $v = 1$ on the electron-doped weak-moiré side and extends across the $n$-$D$ phase diagram toward the hole-doped weak-moiré side. A superconducting resistance dip appears along part of this feature at zero in-plane field (Fig. 1d). Applying an in-plane field of 0.2 T markedly enhances the superconducting response and extends it continuously along the feature (Fig. 1e). As shown in Fig. 1f, increasing the in-plane field initially strengthens the superconducting response and shifts the boundary toward higher carrier density, before suppressing it near 1 T. Similar field dependences are observed across the three superconducting regions, SC1–SC3 (see Fig. S3a-g in Supplementary Note 2). The field-induced boundary shift is consistent with a change in spin polarization across the boundary, in line with recent scanning SQUID measurements on 13-layer rhombohedral graphene [14]. The superconducting response therefore emerges near a putative spin-polarization boundary.

The superconducting resistance dip weakens as the temperature increases from 16 to 84 mK (Fig. 1g). Differential-resistance measurements reveal finite-bias critical-current features, and the superconducting response is rapidly suppressed by a small out-of-plane magnetic field of approximately 25 mT (Inset in Fig. 1g). Defining the superconducting transition temperature $T_c$ at 50% of the normal-state resistance, we obtain $T_c \approx 58$, 35, and 29 mK for SC1, SC2, and SC3, respectively (Fig. 1h). Nevertheless, the resistance remains finite at base temperature, approaching approximately 20 Ω in Fig. 1h, with a minimum $R_{xx}$ of approximately 10 Ω across the superconducting region (see Fig. S3i in Supplementary Note 2).

Finite residual resistance has also been reported in other rhombohedral graphene superconductors [14,16–18]. To investigate its possible origin, we examine the dependence of residual resistance on the gate voltage width of the superconducting region. An inverse correlation is observed when the residual resistance exceeds approximately 60 Ω, whereas no clear correlation is found below this scale (see Fig. S3j-l in Supplementary Note 2). Furthermore, Berezinskii–Kosterlitz–Thouless (BKT) analysis suggests a transition temperature below the base temperature for SC1, and approximately 32 and 17 mK for SC2 and SC3, respectively (see Fig. S4 in Supplementary Note 2). The narrow superconducting region and low BKT transition temperatures may contribute to the finite residual resistance, although its microscopic origin remains unresolved. Moreover, the strong anisotropy between the in-plane and out-of-plane critical fields, together with an in-plane field scale approaching 1 T that substantially exceeds the weak-coupling Pauli limit estimated from $T_c$, is consistent with superconductivity associated with a spin-polarized parent state.

**B. Valley-imbalanced Fermi surfaces**

We first examine superconductivity on the weak-moiré side, where the relevant carriers are predominantly localized on the bottom surface. Figure 2a shows that superconductivity emerges below $D$ = 0.635 V/nm along the putative spin-polarization boundary and then follows a top-gate-voltage trajectory. The layer polarization and electrostatic screening are captured by self-consistent Hartree calculations (Fig. 2b). At large displacement fields, carriers predominantly occupy either the top or bottom surface, whereas both surfaces are occupied in an intermediate diamond-shaped region of the $n$-$D$ plane (see Fig. S5 in Supplementary Note 3). This dual-surface region coincides with the experimentally observed single-gate-tracking regime.

The screening effect is particularly pronounced in thick rhombohedral graphene. As shown in Fig. 2c, along the superconducting trajectory (light-blue dashed line in Fig. 2b), the calculated bottom-surface electron pocket remains nearly unchanged, apart from an abrupt decrease upon entering the dual-surface region. Additional holes introduced by the top gate occupy the top surface or are distributed across the layers, screening the electrostatic potential acting on the bottom-surface carriers. Consequently, the Fermi pocket associated with superconductivity remains relatively insensitive to the top-gate voltage across the superconducting trajectory. To minimize contributions from additional Fermi pockets in the dual-surface region, we first analyze quantum oscillations in the bottom-surface-polarized regime.

To characterize the Fermi surfaces associated with superconductivity, we measure Shubnikov–de Haas (SdH) oscillations (Fig. 2d) at $D$ = 0.618 V/nm, where the zero-field resistance peak evolves into a superconducting dip at $B_{\parallel}$ = 0.2 T (Fig. 2e). The Fermi-surface information is extracted by fast Fourier transform (FFT), with the oscillation frequency $B_f$ related to the carrier density by $B_f = nh/ge$ for $g$ degenerate pockets. Here $h$ is the Planck constant and $e$ is the elementary charge. We define the normalized frequency as $f_\nu = eB_f/(nh)$, which equals $1/g$. Figure 2f shows the SdH oscillations at $n$ = 0.92 × $10^{12}$ /cm$^2$, corresponding to the superconducting region. The FFT spectrum in Fig. 2g reveals two principal frequencies, $f_1$ and $f_2$, on opposite sides of $f_0$ = 1/2 ($B_f \approx$ 19 T), together with a combination peak at $f_1 + f_2 \approx 1$.

Figure 2h shows the evolution of the FFT spectra with carrier density. The splitting of the $f_0$ = 1/2 branch emerges below $n \approx 1.3 \times 10^{12}$ /cm$^2$ and increases with decreasing $n$. Under the spin-polarized assignment, we attribute this splitting to unequal Fermi-surface areas in the K and K′ valleys. The corresponding valley imbalance is quantified by $\Delta n = (f_2 - f_1)n$, reaching $\Delta n = (0.75 \pm 0.1) \times 10^{11}$ /cm$^2$ near superconductivity (Fig. 2i). The uncertainty is dominated by the limited frequency resolution of the FFT spectra. Additional frequency analysis is provided in Supplementary Note 4.

We further examine whether the valley-contrasting orbital Zeeman effect can account for the observed frequency splitting. We simulate quantum oscillations using two parabolic valley bands with an orbital Zeeman splitting $\Delta_B = 2g_{orb}\mu_B B_\perp$ and a zero-field valley splitting $\Delta_0$. Taking $g_{orb} = 10$ and an effective mass $m^* = 0.8m_e$, we find that the orbital Zeeman effect alone ($\Delta_0 = 0$) does not produce a splitting of the principal FFT peak, whereas a finite zero-field splitting of $\Delta_0 = 0.6$ meV generates two distinct frequencies, $f_1$ and $f_2$ (see Fig. S7 in Supplementary Note 3). These results indicate that the observed frequency splitting cannot be explained by the orbital Zeeman effect, supporting a zero-field valley imbalance.

The valley imbalance can be further tuned by the displacement field along the putative spin-polarization boundary. As shown in Fig. 3a, an anomalous Hall response with finite remanence at zero perpendicular field appears at $D > 0.680$ V/nm (see Fig. S2 in Supplementary Note 1 for additional Hall measurements). At $\nu = 1$, the Hall resistance exhibits pronounced hysteresis with an anomalous component of approximately 10 kΩ (Fig. 3b), indicating an incipient Chern insulator. The corresponding FFT spectra of quantum oscillations at $B_\parallel = 0.2$ and 0.6 T are shown in Fig. 3c and 3d, respectively. At high displacement fields ($D > 0.680$ V/nm), the oscillations exhibit a onefold-degenerate sequence, consistent with a spin- and valley-polarized metal and the observed anomalous Hall response.

As $D$ is reduced toward the superconducting region, the onefold-degenerate Fermi pocket evolves into two spin-polarized but valley-imbalanced Fermi pockets. In this regime, no resolvable anomalous Hall response is observed, while the resistance is strongly suppressed by an in-plane magnetic field (Fig. 3e). At $B_\parallel = 0.2$ T, the valley imbalance decreases with decreasing $D$ and remains $\Delta n = (0.75 \pm 0.1) \times 10^{11}$ /cm$^2$ near superconductivity (Fig. 3f). Increasing $B_\parallel$ to 0.6 T strongly suppresses the valley imbalance over $D = 0.65$-$0.68$ V/nm. With further reduction of $D$, the imbalance recovers to a magnitude comparable to that observed at $B_\parallel = 0.2$ T. These results show that the valley imbalance is sensitive to the in-plane magnetic field, particularly near the spin-polarization boundary.

We further examine the evolution of the valley-imbalanced Fermi surfaces as the superconducting trajectory enters the dual-surface region (see Fig. S11 in Supplementary Note 4). Along a fixed bottom-gate trajectory, the split feature near $B_f \approx 20$ T persists over a range of top-gate-induced density ($n_t$), while the oscillations eventually evolve into a single frequency branch, $f_0$. Representative FFT spectra reveal both the split features and the robust $f_0$ branch. The persistence of the oscillation frequency is consistent with the screening mechanism discussed above, in which the bottom-surface Fermi pocket remains relatively insensitive to the top gate as additional carriers occupy the top surface or are distributed across the layers. This raises the possibility of coexisting superconducting and normal states on different surfaces, a scenario that warrants further investigation.

### C. Nonlinear d*V*/d*I* response in an adjacent metallic regime

We next examine the metallic regime adjacent to superconductivity. Figure 4a shows a resistive boundary separating the normal-metal (NM) and spin-polarized-metal (SP-M) regions. Approaching this boundary from the SP-M side, the metallic state develops pronounced nonlinear d$V$/d$I$ features, which evolve into a superconducting response at $B_{\parallel}$ = 0.2 T (Fig. 4b). The superconducting response is initially enhanced by an in-plane magnetic field and is suppressed near $B_{\parallel} \approx 1$ T (Fig. 4c). As shown in Fig. 4d, $R_{xx}(T)$ exhibits a superconducting transition at $B_{\parallel}$ = 0.2 T, whereas a failed superconducting transition is observed at $B_{\parallel}$ = 0 T. At higher temperatures, both traces exhibit similar insulating-like behavior. The superconducting transition temperature is $T_c \approx 95$ mK, while the estimated BKT transition temperature lies below the base temperature (see Fig. S12 in Supplementary Note 5).

The nonlinear metallic state exhibits finite-bias d$V$/d$I$ peaks at currents larger than the superconducting critical-current features (Fig. 4e). The characteristic peak current also increases with increasing $B_{\parallel}$. Figures 4f and 4g compare the temperature dependence of the differential resistance in the superconducting and nonlinear metallic regimes, respectively. The nonlinear response persists up to a characteristic temperature $T^* \approx 293$ mK, substantially higher than $T_c \approx 95$ mK.

Interestingly, the nonlinear d$V$/d$I$ response exhibits slow oscillations under an out-of-plane magnetic field (see Fig. S13 in Supplementary Note 5). The oscillation frequency is approximately 1 T, much lower than the SdH frequency of approximately 18 T. Increasing $B_{\parallel}$ from 0 to 0.2 T further reduces the slow oscillation frequency to approximately 0.5 T (see Fig. S14 in Supplementary Note 5). Two possible origins are considered. One is magneto-intersubband resistance oscillations (MISO) associated with scattering between the two spin-polarized valley subbands with unequal carrier densities [55]. The other is Brown-Zak oscillations arising from charge ordering that breaks translational symmetry. In the latter scenario, an in-plane magnetic field may suppress the competing charge order and favor superconductivity. However, the present measurements do not distinguish between these possibilities.

The origin of the pronounced nonlinear d$V$/d$I$ peaks also remains unresolved. In a charge-ordered state, a dc current may drive the motion of charge textures, whereas in a valley-imbalanced orbital-magnetic state, current-driven orbital-magnetization domain-wall dynamics could produce similar nonlinear features [56]. The continuous evolution of the nonlinear response into the superconducting critical-current features, together with their sensitivity to in-plane magnetic field, suggests a close connection between the nonlinear metallic state and superconductivity. Understanding the nature of this adjacent metallic state may therefore provide further insight into the emergence of superconductivity.

**D. Moiré-selective superconductivity**

Finally, we investigate how the hBN moiré potential affects superconductivity. Fig. 5a compares $R_{xx}$ as a function of $n$ at $D = \pm$ 0.62 V/nm and $B_{\parallel}$ = 0.2 T. A superconducting response is observed only on the weak-moiré side, where carriers are polarized toward the bottom surface and away from the hBN-aligned interface. No superconducting signature is observed on the strong-moiré side. To identify the origin of the Fermi surface on the two sides, we extract the Hall carrier density as shown in Fig. 5b. On the weak-moiré side, $n_H$ follows the reference line $n_H = n$ for $n < 1.8 \times 10^{12}$/cm$^2$, indicating a Fermi surface counted from the charge neutrality point. By contrast, on the strong-moiré side, $n_H$ first follows $n_H = n - n_0/4$ and then crosses over toward $n_H = n - n_0/2$, indicating Fermi surfaces counted from $\nu$ = 1 and $\nu$ = 2 of the moiré miniband, respectively. Band calculations further reveal that the weak-moiré-side band structure remains nearly unchanged upon introducing the moiré potential, whereas the strong-moiré side exhibits substantial band reconstruction and the emergence of an isolated conduction band (Figs. 5d and 5e; see also Fig. S6 in Supplementary Note 3). These Hall density offsets, together with the band calculations, support moiré-induced Fermi-surface reconstruction on the strong-moiré side. The comparison establishes layer-selective superconductivity: it survives when the relevant Fermi surface is remote from the moiré potential but is suppressed on the hBN-aligned surface, where the electronic structure is strongly reconstructed.

## III. DISCUSSION

The in-plane-field enhancement of superconductivity in the valley-imbalanced, spin-polarized metal motivates a phenomenological magnon-assisted pairing scenario (Supplementary Note 3 and Fig. S8). We consider an equal-spin intervalley channel and assume that field-induced spin mixing activates the coupling of quasiparticles to a transverse collective spin mode of the spin-polarized background. Virtual exchange of this mode may provide an effective attraction between quasiparticles on the K and K' Fermi surfaces, offering a possible explanation for the initial field enhancement. A related field-activated mechanism was proposed for Bernal bilayer graphene, in which an in-plane magnetic field enables spin-triplet pairing mediated by soft isospin fluctuations at the onset of isospin order [57].

Other pairing mechanisms cannot be excluded. Acoustic-phonon theories for graphene multilayers support both spin-singlet and spin-triplet intervalley channels [58,59], so a spin-polarized parent state does not by itself identify magnetic fluctuations as the pairing interaction. Intervalley-coherence fluctuations provide another proposed route to triplet pairing in a spin-polarized metal [60].

A potentially complementary explanation involves intrinsic spin–orbit coupling and a field-induced reduction of valley imbalance. The intrinsic Kane–Mele term can be written as $H_I = \lambda_I \tau_z \sigma_z s_z$, where $\tau_z$, $\sigma_z$ and $s_z$ act on valley, sublattice and spin, respectively [61]. Projection onto a surface-dominated band gives an effective valley-dependent Ising term $\lambda_{eff} \tau_z s_z$, with $\lambda_{eff} = \lambda_I \langle \sigma_z \rangle$. If the two valleys have predominantly parallel spins with a finite out-of-plane component, an in-plane field could cant the spins and reduce the spin–orbit contribution to the valley energy offset. If this contribution reinforces the valley imbalance, its reduction could improve Fermi-surface matching for predominantly spin-aligned intervalley pairing. Intrinsic spin–orbit coupling is known to affect the competition between valley-imbalanced and intervalley-coherent states in hBN-encapsulated rhombohedral trilayers [62]. In our device, the field-induced reduction of valley imbalance occurs at $D = 0.65$–$0.68$ V/nm (Fig. 3f), outside the superconducting region. This mechanism therefore remains conditional, and finite-thickness orbital effects of the in-plane field also require consideration.

In summary, we identify a valley-imbalanced, spin-polarized metal as a candidate parent state for superconductivity in thick rhombohedral graphene. The in-plane-field-enhanced superconducting response is reproduced in a second 15-layer device (D2), as detailed in Supplementary Note 6. The spin polarization permits an equal-spin intervalley pairing channel, whereas the unequal K and K' pocket areas constrain conventional zero-momentum intervalley pairing. The field enhancement is consistent with a phenomenological magnon-assisted scenario. This system thus provides an electrically tunable platform for exploring unconventional pairing through control of spin, valley and layer polarization.

## Data availability

The data generated in this study are provided in the Source Data file.

**Acknowledgments**

This work is supported by A*STAR (M24M8b0004), Singapore National Research Foundation (NRF-CRP30-2023-0003, NRF-CRP31-0001, NRF2023-ITC004-001 and NRF-MSG-2023-0002) and Singapore Ministry of Education Tier 2 Grant (MOE-T2EP50222-0018).

**Author contributions**

W.G. supervised the project. L.L. and J.Z. fabricated the devices and performed the electrical-transport measurements. C.X., Y.Z., and L.L. performed theoretical calculations. K.W. and T.T. provided hexagonal boron nitride crystals. L.L., J.Z., Y.Z., C.X., and W.G. analyzed the data. L.L., Y.Z., C.X., J.Z., and W.G. wrote the paper with the input from all the authors.

**Competing interests**

The authors declare no competing interests.

## Figure & Figure captions

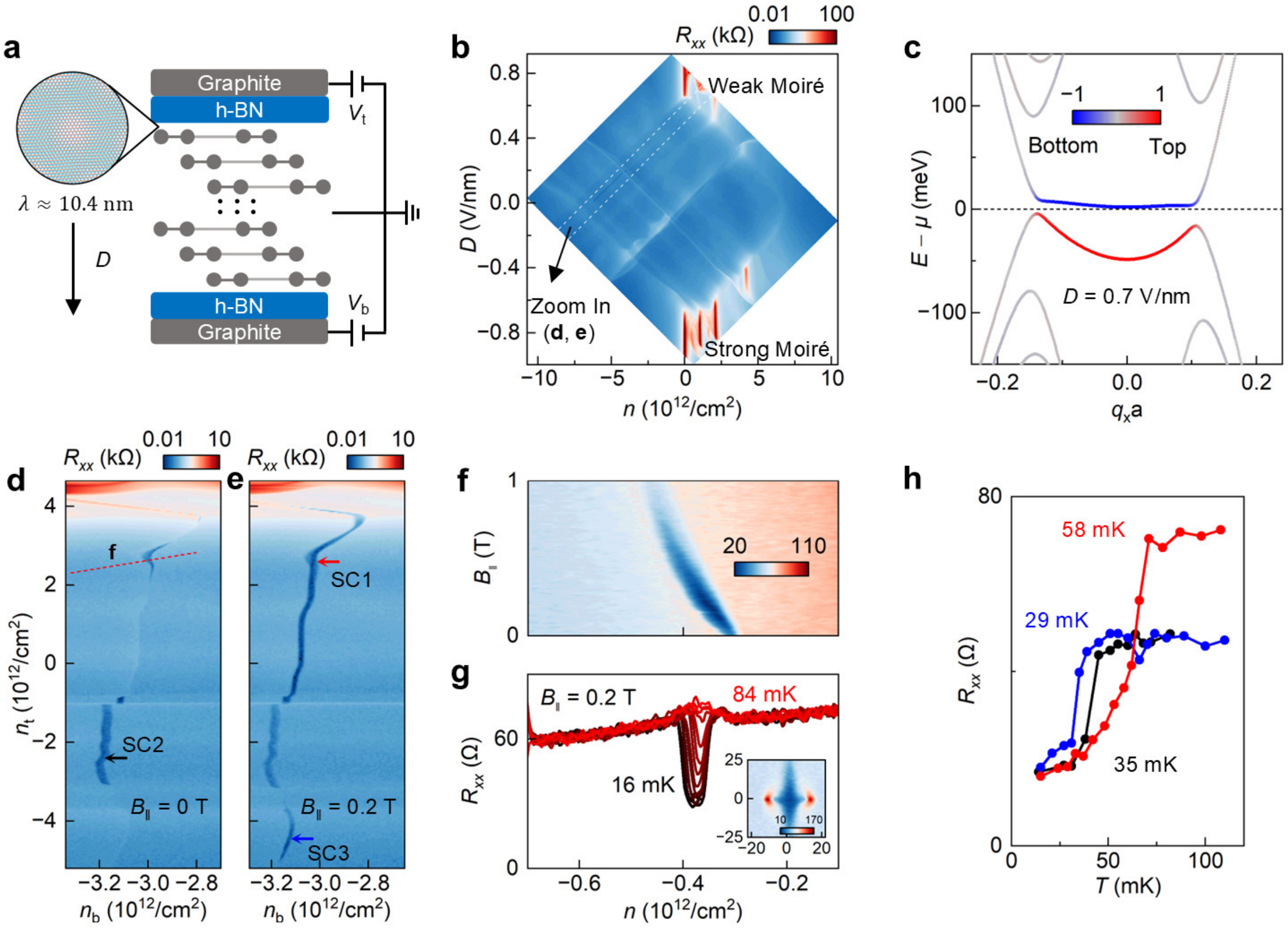

**Fig. 1. In-plane-field-enhanced superconductivity.** **a**, Schematic of the dual-gated 17-layer rhombohedral graphene device. Rhombohedral graphene is aligned with the top hBN, forming a moiré pattern with a period of 10.4 nm. **b**, Longitudinal resistance $R_{xx}$ as a function of carrier density $n$ and displacement field $D$. **c**, Calculated band structure of 17-layer rhombohedral graphene at $D = 0.7$ V/nm. A flat conduction band edge is resolved, with the corresponding low-energy states predominantly localized on the bottom graphene surface. The black dashed line denotes the chemical potential. **d**, **e**, $R_{xx}$ as a function of $n_b$ and $n_t$ at $B_\parallel = 0$ T and 0.2 T, respectively. The plotted region corresponds to the area enclosed by white dashed lines in **b**. Arrows mark SC1–SC3. **f**, $R_{xx}$ as a function of $n$ and $B_\parallel$, measured along the red dashed line in **d**. **g**, Linecuts of $R_{xx}$ versus $n$ at $B_\parallel = 0.2$ T, measured from 16 mK to 84 mK. Inset, d$V$/d$I$ map as a function of $I_{dc}$ (nA) and $B_\perp$ (mT). **h**, Temperature dependence of $R_{xx}$ in the superconducting region of SC1-SC3.

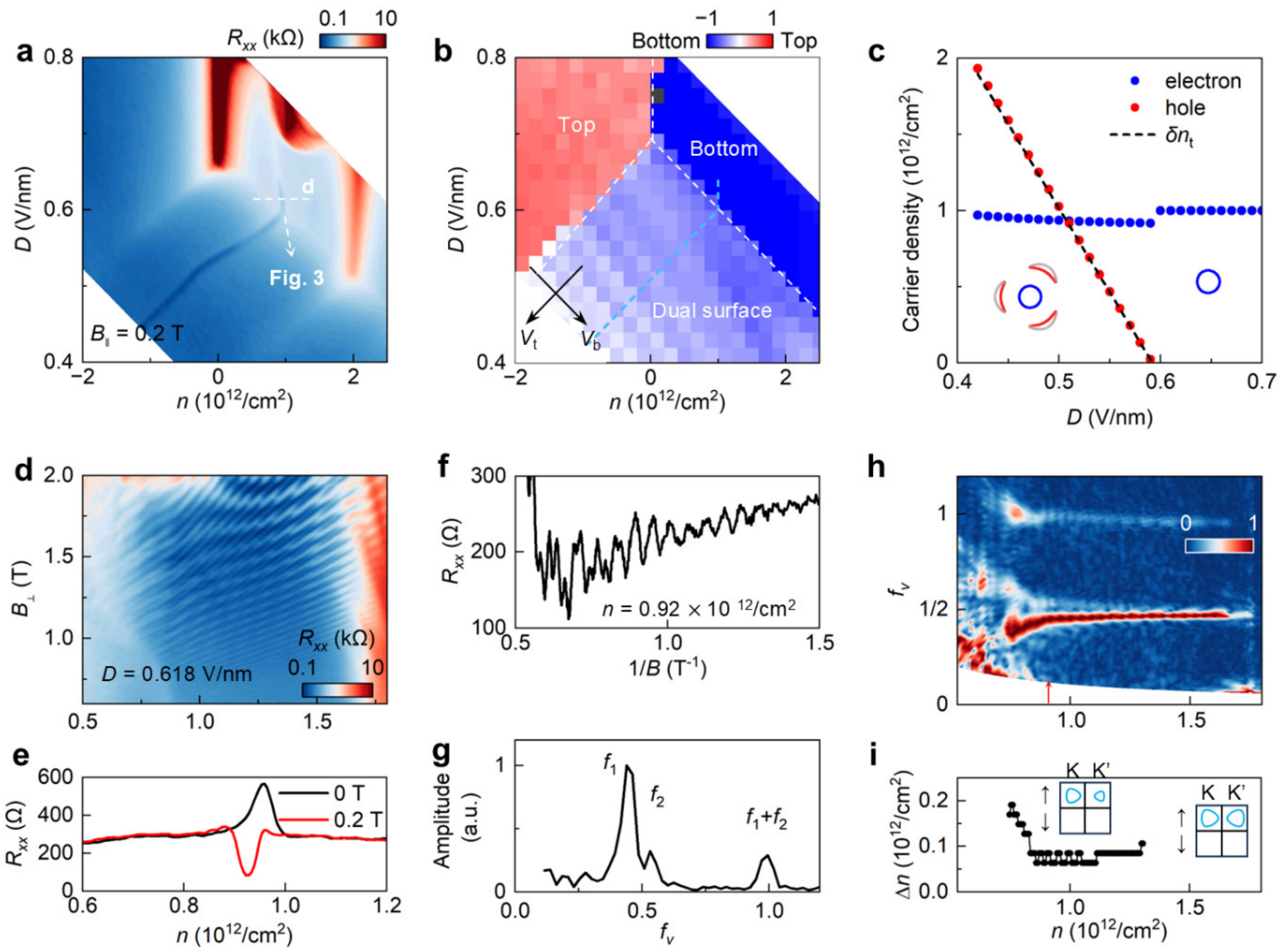

**Fig. 2. Valley-imbalanced Fermi surfaces.** **a**, Zoomed-in $R_{xx}$ ($n$, $D$) at $B_{\parallel}$ = 0.2 T on the weak-moiré side. **b**, Calculated layer polarization as a function of $n$ and $D$. The light blue dashed line follows the superconducting region. **c**, Calculated carrier densities of Fermi pockets versus $D$ along the light blue dashed line in **b**. The black dashed line indicates the change in top-gate-doped density. The left Fermi pockets are calculated at $D$ = 0.5 V/nm, the right at $D$ = 0.62 V/nm. **d**, $R_{xx}$ as a function of $n$ and $B_{\perp}$ at $D$ = 0.618 V/nm. **e**, $R_{xx}$ versus $n$ at $D$ = 0.618 V/nm with $B_{\parallel}$ = 0 and 0.2 T. **f**, $R_{xx}$ versus $1/B_{\perp}$ extracted from **d** at $n = 0.92 \times 10^{12}$/cm$^2$. **g**, FFT spectrum of **f**. Three frequency components are marked as $f_1$, $f_2$, and $f_1 + f_2$. **h**, FFT spectra of **d**. The red arrow indicates the carrier density at which the superconductivity is observed. **i**, Valley imbalance $\Delta n$ versus $n$. The inset shows Fermi pockets in four sectors with spin and valley. The valley imbalance develops as the carrier density approaches the superconducting region.

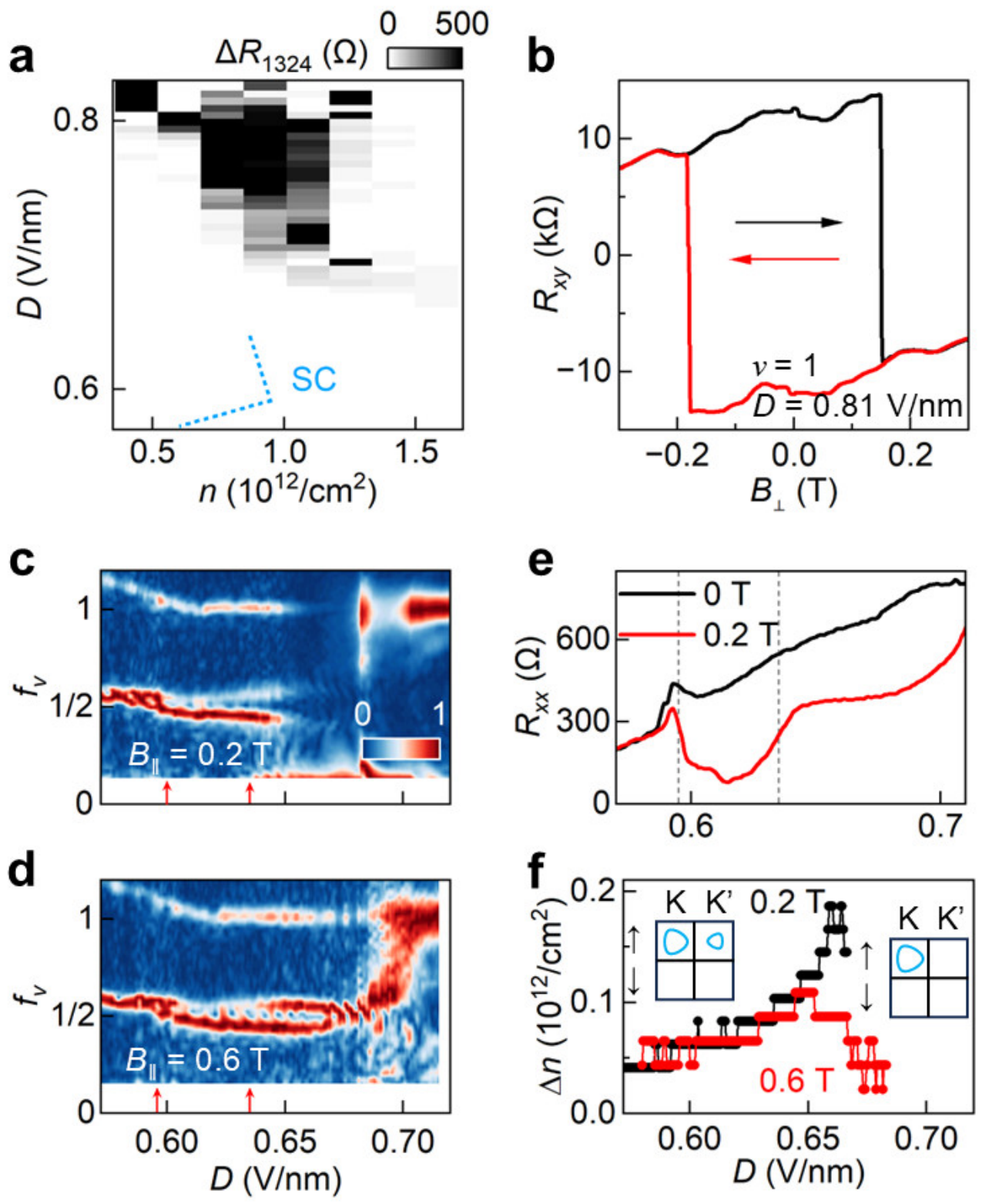

**Fig. 3. Displacement field dependence of valley imbalance. a,** Zero-field hysteresis amplitude $\Delta R_{1324}$ as a function of $n$ and $D$. $\Delta R_{1324} = |\Delta R_{1324}(+) - \Delta R_{1324}(-)|$. Here (+) and (–) correspond to the sweep direction of magnetic field. The light-blue dashed line follows the superconducting region. **b,** $R_{xy}$ versus $B_{\perp}$ at $\nu = 1$, showing an anomalous Hall response. Arrows indicate sweep directions. **c**, **d**, FFT spectra of $R_{xx}$ ($D$, $B_{\perp}$) at $B_{\parallel} = 0.2$ and 0.6 T, respectively. SdH oscillations are measured along the trajectory indicated by the white dashed arrow in Fig. 2a. **e**, $R_{xx}$ as a function of $D$ at $B_{\parallel} = 0$ and 0.2 T, measured along the same trajectory with **c** and **d**. **f**, Valley imbalance $\Delta n$ versus $D$. The inset shows Fermi pockets in four sectors with spin and valley.

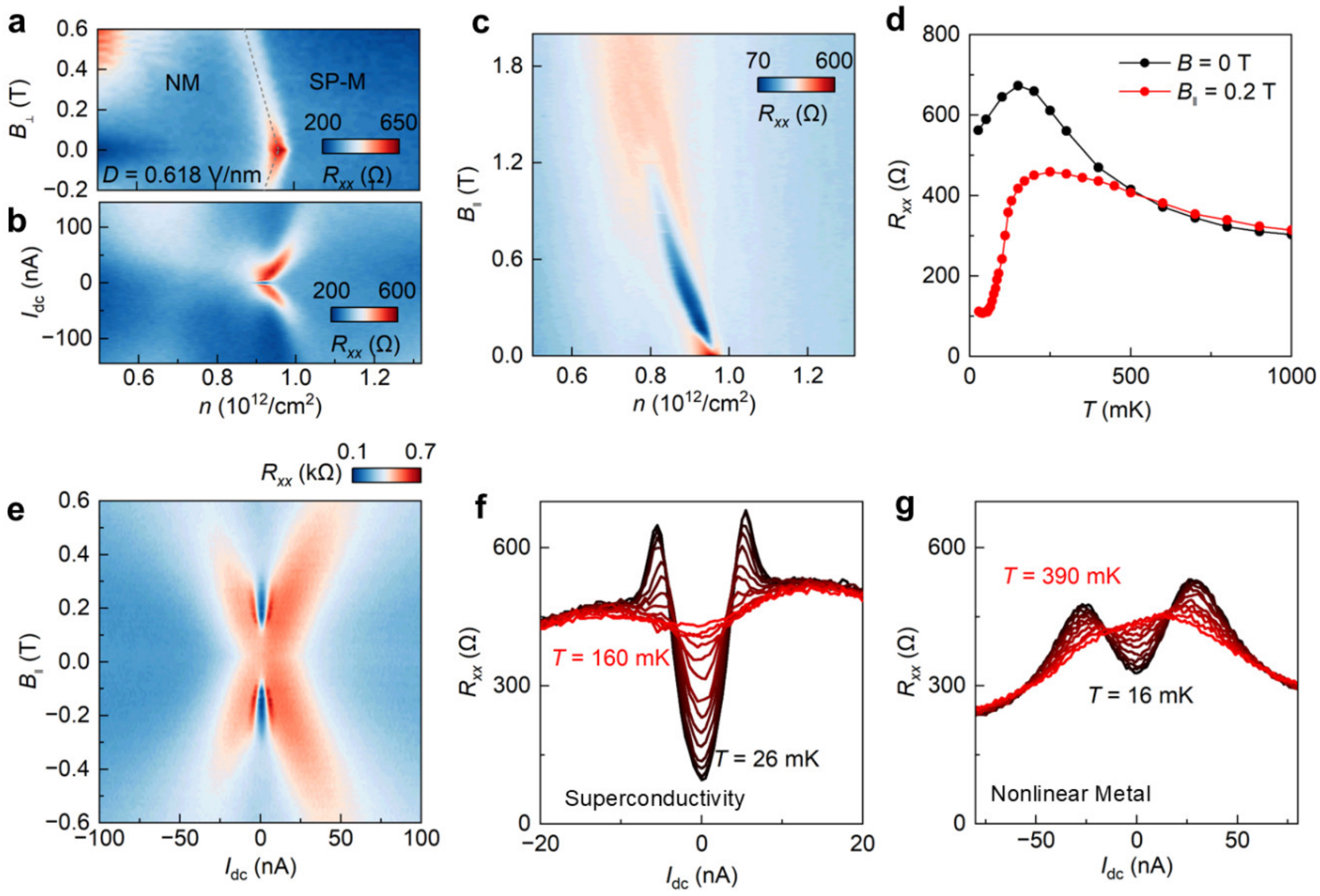

**Fig. 4. Nonlinear metallic response adjacent to superconductivity.** **a**, $R_{xx}$ as a function of $n$ and $B_\perp$ at $D = 0.618$ V/nm and $B_\parallel = 0$. **b**, d$V$/d$I$ as a function of $n$ and $I_{dc}$ at $D = 0.618$ V/nm and $B_\parallel = 0.2$ T. **c**, $R_{xx}$ as a function of $n$ and $B_\parallel$ at $D = 0.618$ V/nm and $B_\perp = 0$. **d**, $R_{xx}$ versus $T$ at $B_\parallel = 0$ and $B_\parallel = 0.2$ T, measured at the phase boundary in **a**. **e**, d$V$/d$I$ as a function of $I_{dc}$ and $B_\parallel$ at the phase boundary. **f**, d$V$/d$I$ versus $I_{dc}$, measured in the superconducting region. **g**, d$V$/d$I$ versus $I_{dc}$, measured in the nonlinear metallic region.

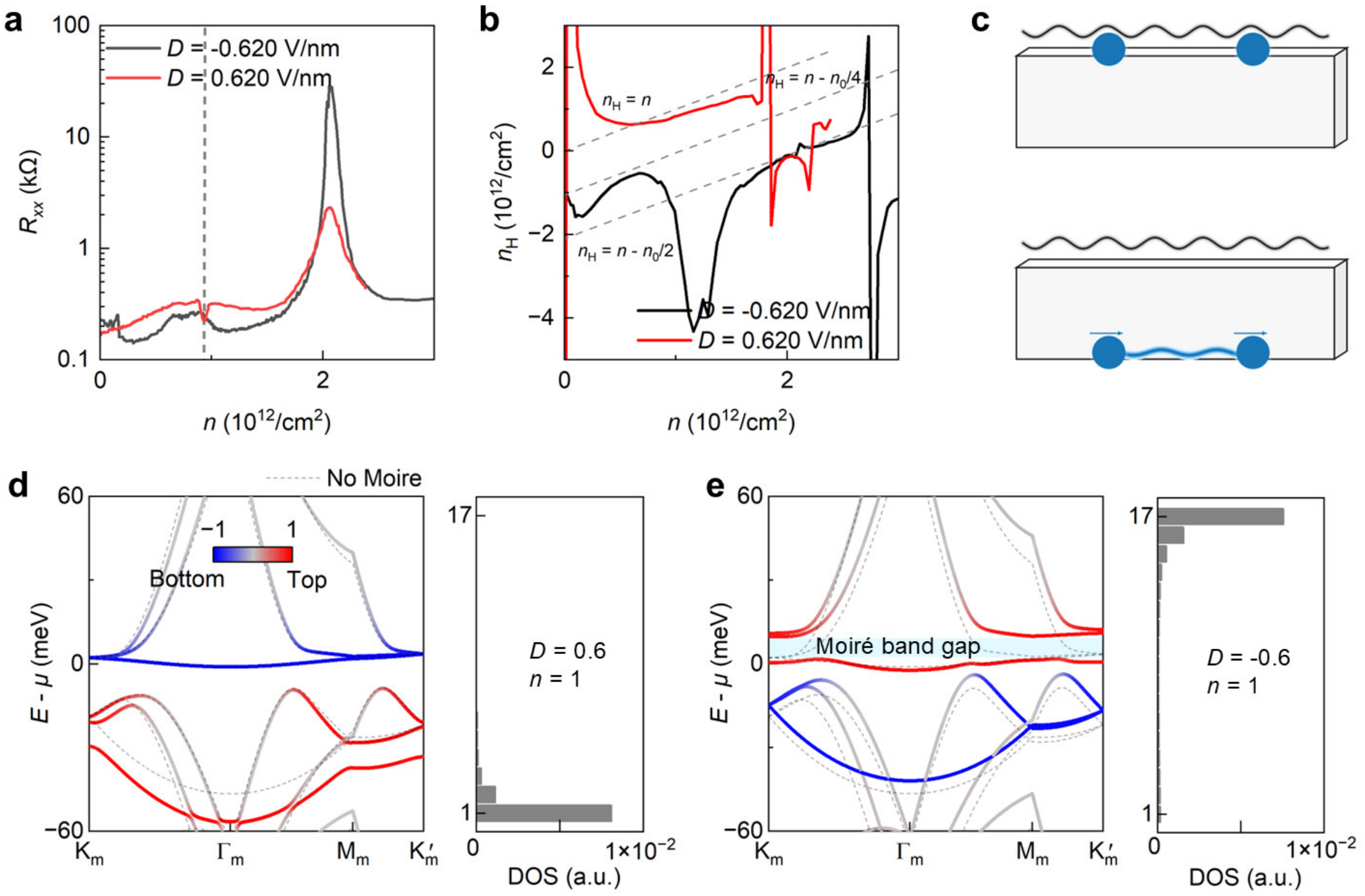

**Fig. 5. Moiré-selective superconductivity. a**, $R_{xx}$ versus $n$ at $D = \pm0.620$ V/nm and $B_{\parallel} = 0.2$ T. **b**, Hall carrier density $n_H$ versus $n$ at $D = \pm0.620$ V/nm. The gray dashed lines $n_H = n$, $n_H = n - n_0/4$, and $n_H = n - n_0/2$ are reference lines, corresponding to Fermi pockets counted from charge neutrality, $v = 1$, and $v = 2$, respectively. Here $n_0$ is the full-filling carrier density. **c**, Schematic of pairing conditions on the strong- and weak-moiré sides. Pairing is suppressed on the strong-moiré side owing to Fermi-surface reconstruction, whereas it survives on the weak-moiré side. **d**, **e**, Left, calculated band structures on the weak-moiré side ($D = 0.6$ V/nm and $n = 1 \times 10^{12}$ /cm$^2$) or the strong-moiré side ($D = -0.6$ V/nm and $n = 1 \times 10^{12}$ /cm$^2$). The gray dashed curves show the band structure without a moiré potential. Right, calculated layer-resolved DOS. Layers 1-17 are indexed from the bottom to the top graphene surface.